\documentclass[aps,pra,twocolumn]{revtex4}
\usepackage{}
\usepackage{amsfonts}
\usepackage{amssymb}
\usepackage{color}
\usepackage{amsmath}
\usepackage{graphicx}
\usepackage{subfigure}
\usepackage{txfonts}
\usepackage{bm}
\usepackage{ulem}

\newcommand{\scamat}{\mathbb{S}}
\newcommand{\identity}{\mathbb{I}}

\begin{document}

\title{Information scrambling in a collision model}

\author{Yan Li, Xingli Li, and Jiasen Jin}
\email{jsjin@dlut.edu.cn}
\affiliation{School of Physics, Dalian University of Technology, Dalian 116024, China}

\begin{abstract}
The information scrambling in many-body systems is closely related to quantum chaotic dynamics, complexity, and gravity. Here
we propose a collision model to simulate the information dynamics in an all-optical system. In our model the information is initially localized in the memory and evolves under the combined actions of many-body interactions and dissipation.
We find that the information is scrambled if the memory and environmental particles are alternatively squeezed along two directions which are perpendicular to each other. Moreover, the disorder and imperfection of the interaction strength tend to prevent the information flow away to the environment and lead to the information scrambling in the memory. We analyze the spatial distributions of the correlations in the memory. Our proposal is possible to realize with current experimental techniques.
\end{abstract}
\date{\today}
\maketitle

\section{Introduction}

In quantum mechanics, the evolution of a closed system is described by a unitary transformation. The encoded information in the system is thus preserved within the system during the time evolution because of the unitarity of the transformation. Although the information about the initial state will not be erased, it may spread throughout the entire system and cannot be accessed by local measurements. The delocalization of the initially localized information in a quantum system is referred to as {\it information scrambling}. Information scrambling is ubiquitous in a variety physical systems ranging from the black hole in gravity to the many-body system in the field of condensed matter. It is intimately related to the phenomena of entanglement propagation in a diffusive system \cite{kim2013,khemani2018}, transportation in non-Fermi liquids \cite{banerjee2017}, local thermalization in non-equilibrium many-body systems \cite{nandkishore2015,alba2019}, and the black-hole information paradox \cite{hayden2007}. Information scrambling can also be considered the reverse process of information encoding in a neural networks \cite{shen2019}.

In particular, the information scrambling can be employed as an indicator of the quantum chaotic dynamics. In the Heisenberg picture, a local unitary operator $\hat{W}$, which can be considered as the locally encoded information, of a quantum many-body system with Hamiltonian $\hat{H}$ will evolve as $\hat{W}(t)=e^{i\hat{H}t}\hat{W}e^{-i\hat{H}t}$. Depending on the many-body Hamiltonian $\hat{H}$, the operator $\hat{W}(t)$ may spread over the entire system and become nonlocal. The delocalization of the initial-local operator is recognized as the quantum version of butterfly effect and is essentially the same as the information scrambling. An appropriate quantity to characterize such a process is the so-called out-of-time-order correlator (OTOC) $C(t)=\langle\hat{W}^\dagger(t)\hat{V}^\dagger\hat{W}(t)\hat{V}\rangle$ \cite{swingle2018} where $\hat{V}$ is another local unitary operator which does not overlap with $\hat{W}$ at initial time, i.e. $[\hat{W}(0),\hat{V}]=0$.  As the time goes, the commutator $[\hat{W}(t),\hat{V}]$ becomes nonzero implying the nonlocality of $\hat{W}(t)$. The non-community of $\hat{W}(t)$ and $\hat{V}$ is related to the OTOC by $\langle|[\hat{W}(t),\hat{V}]|^2\rangle=2(1-\text{Re}[C(t)])$. Therefore the decay of OTOC means the occurrence of information scrambling.
Notice that the terms $e^{-i\hat{H}t}$ and $e^{i\hat{H}t}$ in the expression of $\hat{W}(t)$ are the forward and backward time-evolution operator. The latter is crucial in defining the OTOC. Although recent experimental works have realized the time-reversal operation in the nuclear magnetic resonance quantum simulator \cite{li2017} and trapped-ion system \cite{garttner2017}, it is still challenging in measuring the OTOC in large-scale realistic systems.

An alternative measure of information scrambling is the tripartite mutual information (TMI). Usually, the negative TMI during the time evolution means that the information is scrambled. It has been proven that the TMI is essentially equivalent to the OTOC in capturing the feature of information scrambling by means of the channel-state duality \cite{hosur2016}. Studies based on the TMI-measured information scrambling in the spin XXX model and the Sachdev-Ye-Kitaev model show that the scrambling is an independent property of the integrability of the Hamiltonian \cite{iyoda2018}.

So far, in most of the studies the exact diagonalization of the many-body Hamiltonian are commonly used to compute the OTOC or TMI \cite{iyoda2018,shen2017,pappalardi2018}. However, the numerical approaches are limited by the dimensions of the Hilbert space of the many-body system. Here we utilize the {\it collision model} (CM) to simulate the process of quantum information scrambling directly. The CM was firstly proposed to simulate the Markovian dynamics of open quantum systems in a stroboscopic way \cite{scarani2002}. In the CM, the system is represented by a particle and the environment is represented by an ensemble of uncorrelated identical environmental particles. The interactions between system and environment are simulated by a sequences of collisions of system and environment particles. By introducing intra-collisions between environment particles or appropriately embedding quantum correlation in the environment particles, the non-Markovianity may arise \cite{ciccarello2013a,ciccarello2013b,rybar2012,mccloskey2014,bernardes2014,jin2015,cakmak2017,mascarenhas2017,wang2017,jin2018,campbell2018,campbell2019,cuevas2019}. Recently, the CMs are also employed to investigate the thermodynamics properties \cite{strasberg2017,li2018,cusumano2018,man2019a,man2019b,rodrigues2019}, quantum synchronization \cite{karpat2019}, quantum friction \cite{grimmer2019} and multipartite entanglement generation \cite{cakmak2019}.

In this paper, we present a scheme based on the CM to simulate the information scrambling of a continuous variable system. In the scheme, we encode the information of the system into a memory which is contacted to the environment. A realization of our CM in an all-optical interferometric network is also presented. The system, memory and environment are represented by the optical modes and the interactions, or say, the collisions in CM, are implemented on the linear optical elements. We restrict the discussions to the Gaussian state and adopt the TMI as the measure of information scrambling. The model presented in this paper enables us to study the dynamics of information over a large-size many-body memory which is usually difficult to tackle due to the huge dimension of Hilbert space.  So far, the information scrambling in a dissipative memory are discussed only in a few work \cite{syzranov2018,zhang2019}. Here the presented model is able to simulate not only the dynamics of information over the memory but also the leaking of information from the memory to the environment.
We investigate how does the local information in the memory evolve and be scrambled in the stroboscopic evolution due to combined actions of the many-body interaction, disorders and dissipation.

The paper is organized as follows. In Sec. \ref{model}, we explain the idea and the setup of the CM. The mathematical descriptions for the discrete time dynamics of the information basing on the von Neumann entropy are discussed as well. In Sec. \ref{resultsanddiscussion}, we show the dynamics of the information with different states of the memory and environment and investigate the effects of disorder and imperfection of interactions on the information scrambling. We also consider the case of a large size memory through which we study the correlations of the system and the memory particles. We summarize in Sec. \ref{summary}.

\section{A collision model based optical scheme}
\label{model}
In a general CM, the degrees of freedom of the system and the environment are represented by a set of particles which is the unit cell of the model. The arbitrary two-body interaction is simulated by the collision between the corresponding particles. The continuous dynamics of the system can thus be stroboscopically represented via a series of collisions of the system and environmental particles \cite{scarani2002,ciccarello2013a}. Here we adopt the CM to investigate the dynamics of the information in the presence of both many-body interaction and the dissipation.

\subsection{Collision model}\label{Scheme}
The CM we are considering consists of three blocks: the system $S$, the memory $M$, and the environment $E$. As shown in Fig. \ref{s1}, the system is represented by a single particle while the memory are represented by $N_m$ ($N_m\ge 2$) particles. The number of particles in the environmental block increases as the time goes.

At the initial time, the information carried by the system is locally encoded in the memory by entangling the system particle and the first, named $m_1$ particle in the memory. Once the information is encoded, the system mode $S$ is isolated and do not evolve any more. Due to the intra-collisions of the memory particles and the collisions between the memory and its environment, the encoded information will spread over the memory and leak to the environment. The former collisions simulate the many-body interactions in the memory while the latter simulate the dissipation induced by the memory-environmental interaction. We are interested in the time-evolution of the information in the memory, which is driven by the internal collisions as well as the external collisions with the environments.

At each step, the collisions can be divided into three segments as shown in Fig.\ref{s1}: firstly the bipartite collisions take place sequentially between the neighbouring particles in the environmental block. Secondly, the collision between the $e_1$ and $m_{N_m}$ particles takes place. Such a memory-environment collision opens the channel for the information flowing from the memory to the environment. Thirdly, the bipartite collisions take places sequentially between the neighbouring particles in the memory block. The collisions in the three segments can be considered as dynamical maps and denoted by the superoperators $\mathcal{U}_{E}$, $\mathcal{U}_{ME}$ and $\mathcal{U}_{M}$, respectively. Thus the joint state of the total system at step $L$ is given by the following iteration expression, for $L\ge 1$,
\begin{equation}
\rho(L) = \left[ \mathcal{U}_{M}\circ\mathcal{U}_{ME}\circ\mathcal{U}_{E}\right] \rho(L-1),
\label{rhoL}
\end{equation}
where $\rho(L)$ denotes density matrix of the joint system at the end of the $L$-th step and ``$\circ$" denotes the composition of superoperators.

In order to investigate the dynamics of the information which is initially shared by the $S$ and $m_1$ particles, we employ the bipartite mutual information (BMI) and TMI of the subsystem composed of the system and the memory to characterize the information. To be more detailed, the TMI is defined by
\begin{equation}
I_3(S:M_1:M_2) = I_2(S:M_1) + I_2(S:M_2) - I_2(S:M),
\label{eq_TMI}
\end{equation}
here we have already divided the memory block $M$ into two parts $M_1$ and $M_2$. The $M_1$ part contains only the $m_1$ particle while the $M_2$ part contains all the particles left in the memory. The terms in the r.h.s. of Eq. (\ref{eq_TMI}) are the BMIs between the corresponding subsystem. The BMI measures the total correlation between two subsystems of a composite system, and the expression of BMI in Eq. (\ref{eq_TMI}) is defined by
\begin{equation}
I_2(S:X) = S(\rho_S) + S(\rho_X) - S(\rho_{SX}),
\label{eq_BMI}
\end{equation}
with $X = M_1$, $M_2$, and $M$, $\rho_X$ is the corresponding reduced density matrix and $S(\rho)=-\text{tr}\left(\rho\log{\rho}\right)$ is the von Neumann entropy.

The negative TMI, implying $ I_2(S:M)>I_2(S:M_1) + I_2(S:M_2)$, means that the total correlation shared by the system and the whole memory cannot be fully characterized by the sum of the correlations those shared individually by the system and each partition of the memory; some of the information are hidden nonlocally over the memory. In other words, the local measurements on $M_1$ and $M_2$ are not able to reconstruct the information about $S$. Since initially the total correlation is restricted inside $S$ and $M_1$, the negative TMI can be considered as a diagnostic of the information scrambling.

\subsection{CM in an optical interferometer}
\begin{figure}[h]
  \includegraphics[width=0.95\linewidth]{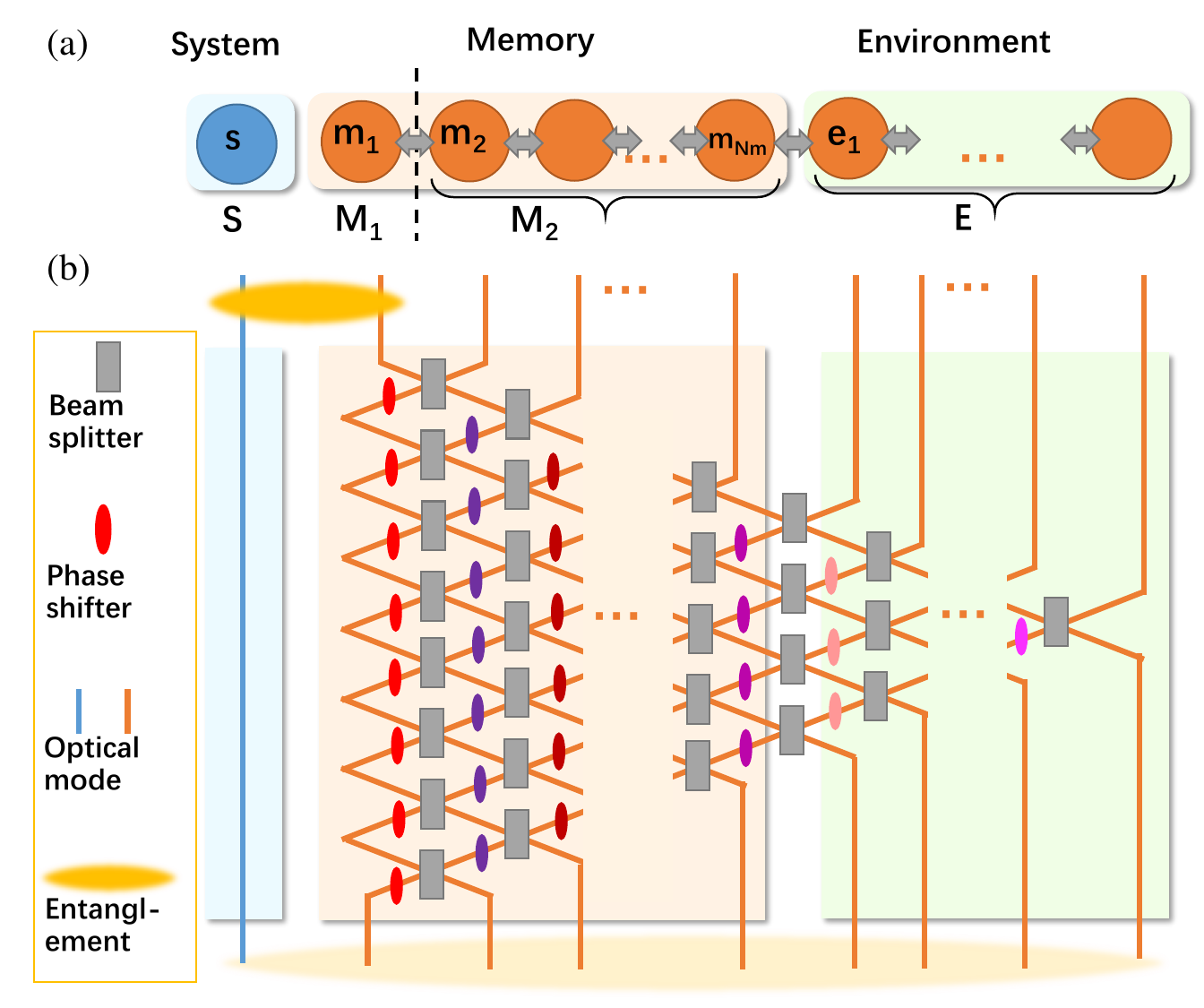}
  \caption{(a) Pictorial illustration of the CM. The model consists of the system, memory and environment blocks which are represented by an ensemble of particles. Moreover, the memory block is divided into two parts: $M_1$ and $M_2$, separated by the vertical dashed line. $M_1$ is the first particle and $M_2$ are the left particles of the memory block. Initially, the information carried by the system particle is locally encoded with the first site of the memory block through entanglement. The memory is in turn in touch with an environment. The information will spread over the memory block and has the possibility to flow out into the environment due to the inter-collision between the system and environment blocks and the intra-collisions in the memory and environment blocks, respectively. (b) The interferometric network to simulate the CM. The network are composed of BSs and phase shifters. The particles of the CM are represented by the optical modes. The mixing of two modes simulates the collisions of the corresponding particles. The phase shifters introduce the disorder.}
  \label{s1}
\end{figure}

In Eq. (\ref{rhoL}), the superoperators $\mathcal{U}_E$, $\mathcal{U}_M$, and $\mathcal{U}_{ME}$ are actually unitary transformations acting on the corresponding particles and they depend on the concrete interactions of the realistic system. Here we discuss the realization of CM in a linear optical system. The unit in the CM is now represented by the optical mode propagates along different optical paths. The collision between two units of CM is simulated by the mixing of two optical modes on a beamsplitter (BS). We remind that the BS transfers two input modes $\hat{a}^{\text{in}}_1$ and $\hat{a}^{\text{in}}_2$ into two output modes $\hat{a}^{\text{out}}_1$ and $\hat{a}^{\text{out}}_2$ through ${\bf \hat{a}}^{\text{out}} = \mathbb{S}_{\text{BS}}{\bf \hat{a}}^{\text{in}}$, where ${\bf \hat{a}}^{\text{in(out)}} = \left[\hat{a}^{\text{in(out)}}_1,\hat{a}^{\text{in(out)}}_2 \right]^{T}$ and $\mathbb{S}_{\text{BS}}$ is the scattering matrix given by
\begin{equation}
\mathbb{S}_{\text{BS}}=
\begin{pmatrix}
 r & t \\
 -t & r \\
\end{pmatrix},
\label{Scattering matrix1}
\end{equation}
with $r=\sin\eta$ and $t=\cos\eta$ being the reflectivity and transmissivity of the BS satisfying $r^2 + t^2 = 1$ and $\eta\in[0,\pi/2]$. The all-optical setup of our CM is illustrated in Fig. \ref{s1}(b). The BSs in the setup constitute an interferometric network. In this work, we restrict that all the input modes are prepared in the Gaussian state such as the vacuum state, the squeezing state and so on. Because the elements composed of the interferometric network are quadratic, the Gaussianity of the state is preserved during the whole stroboscopic time evolution.

Let us start by introducing the states of the input modes. The system mode and the first mode of the memory block are prepared in a two-mode squeezed vacuum (TMSV) state. We recall that the TMSV state is generated by squeezing the two-mode vacuum state $|\xi_{\text{TMSV}}\rangle=\hat{S}(\xi)|0_S0_{M_1}\rangle$,  $\hat{S}(\xi)$ is the two-mode squeezing operator which is defined as
\begin{equation}
\hat{S}(\xi)=\exp{\left(\frac{1}{2}\xi^* \hat{a}_S\hat{a}_{M_1}-\frac{1}{2}\xi \hat{a}_S^\dagger \hat{a}_{M_1}^\dagger\right)},
\end{equation}
where $\hat{a}_S$ and $\hat{a}_{M_1}$ are the annihilation operators of the mode $S$ and $M_{1}$, respectively, and $\xi$ is squeezing parameter and set to be real in this work. The entanglement of the state $|\xi_{\text{TMSV}}\rangle$ measured by the logarithmic negativity is $\xi$ \cite{vidal2002}. The joint input state including all modes can be expressed as the following (the subscript `$J$' stands for the joint state hereinafter),
\begin{equation}
\rho_J^{\text{in}}=|\xi_{\text{TMSV}}\rangle_{S,M_1}\langle\xi_{\text{TMSV}}|\otimes\rho_{M_2}\otimes\rho_E.
\end{equation}
where $\rho_{M_2}$ and $\rho_E$ represent the Gaussian states of the $M_2$ part of the memory and the environment respectively, and will be specified later.

In order to solve the stroboscopic evolution of input state $\rho_J^{\text{in}}$ one needs to compute the scattering matrix $\scamat(L)$ of the whole interferometer at step $L$. We note that $\scamat(L)$ is a $(L+2)\times (L+2)$ square matrix.
It is straightforward to construct the scattering matrix at step $L$ ($L\ge1$) as the following,
\begin{equation}
\label{SL}
\scamat(L) = \left(\prod_{k=1}^{L} \scamat_{k,k+1}\right) \scamat(L-1),
\end{equation}
where the matrices in the cumulative product are in the ascending order of $k$ from right to left and $\scamat_{k,k+1}$ is given by
\begin{equation}
\scamat_{k,k+1} =
\begin{pmatrix}
\identity_{k}&0&0&0\\
0&re^{i\delta_{k}}&te^{i\delta_{k}}&0\\
0&-t&r&0\\
0&0&0&\identity_{L-k}\\
\end{pmatrix}
,
\end{equation}
in which $\identity_{k}$ is the $k\times k$ identity matrix and $\delta_{k}$ is the phase shift. We suppose that all the BSs are  identical and characterized by the transmission angle $\eta$, i.e. $r=\sin\eta$ and $t=\cos\eta$. Therefore, the properties of the scattering matrix are completely determined by the reflectivities and transmissivities of the BSs and the phase shifters. We note that the scattering matrix for $L=0$ is an identity matrix.

\subsection{The covariance matrix and tripartite mutual information}
Since the Gaussianity of the state is preserved during the evolution, it is convenient to describe the state in the characteristic function formalism \cite{walls1994}. The symmetrically ordered characteristic function of $\rho^{\text{in}}_J$ is given by
\begin{equation}
\chi_J^{\text{in}}(\bm{\mu})=\text{tr}\left[\hat{D}(\bm{\mu})\rho_J^{\text{in}}\right],
\label{chi_J}
\end{equation}
where $\hat{D}(\bm{\mu})$ is the multi-mode Weyl displacement operator defined as $\hat{D}(\bm{\mu})=\bigotimes_j\hat{D}(\mu_j)$ with $\bm{\mu}=[\mu_S,\mu_{M_1},...]^T$, of which $\hat{D}(\mu_j)=\exp{\left(\mu_j\hat{a}_j^{\dagger}-\mu_j^*\hat{a}_j\right)}$ is the displacement operator for the $j$-th mode. The interferometric network characterized by the scattering matrix $\scamat$ maps the $\chi_J^{\text{in}}(\bm{\mu})$ to the output characteristic function via the following formula \cite{jin2018}
\begin{equation}
\chi_J^{\text{out}}(\bm{\mu})=\chi_J^{\text{in}}(\scamat^{-1}\bm{\mu}).
\label{chi_Jout}
\end{equation}

On the other hand, the von Neumann entropy of a single-mode Gaussian state $\rho$ is given as the following,
\begin{equation}
S(\rho)=\sum_{k=1}^N{f(\nu_k)},
\end{equation}
where $f(x)=\left(x+\frac{1}{2}\right)\ln{\left(x+\frac{1}{2}\right)}-\left(x-\frac{1}{2}\right)\ln{\left(x-\frac{1}{2}\right)}$ and $\nu_k$ are the symplectic eigenvalues of the so-called covariance matrix \cite{serafini2004,serafini2006}. The single-mode covariance matrix $\sigma$ is the second moment of the characteristic function and its elements are defined by
\begin{equation}
\sigma(j,k) =
\frac{1}{2} \langle \hat{x}_{j}\hat{x}_{k} + \hat{x}_{k}\hat{x}_{j} \rangle - \langle\hat{x}_{j}\rangle \langle\hat{x}_{k}\rangle, (j,k= 1,2),
\label{eq.11}
\end{equation}
 where $\langle\cdot\rangle$ is the expectation value, $\hat{x}_{1} = (\hat{a}_{k} + \hat{a}_{k}^{\dagger})/\sqrt{2}$ and $\hat{x}_{2} = (\hat{a}_{k} - \hat{a}_{k}^{\dagger})/\sqrt{2}i$.
 The symmetrically ordered moments can be obtained through the single-mode characteristic function,
 \begin{equation}
\text{tr}\Big\{\rho\left[(a_{k}^{\dagger})^{p}a_{l}^{q}\right]_{\text{symm}}\Big\}= (-1)^{q}\frac{\partial^{p+q}}{\partial \mu_{k}^{p} \partial \mu_{l}^{\ast q}}\chi(\mu)\Big\vert_{\mu=0} .
 \label{eq.12}
 \end{equation}

Generally, the covariance matrix for an $m$-mode Gaussian state is $2m$-dimensional \cite{wang2017}.

Recall that the characteristic function of the reduced density matrix associated with the $k$-th mode is given by the partial trace of the joint characteristic function over all the modes other than $k$ \cite{wang2007},
\begin{equation}
\chi_k^{\text{out}}(\mu_k)=\chi_J^{\text{out}}(\bm{\mu})\Big\vert_{\bm{\mu}=[0,...,\mu_k,0,...]^T}.
\label{chi_k}
\end{equation}

With the help of Eqs. (\ref{chi_Jout}), (\ref{eq.12}) and (\ref{chi_k}), we can obtain the covariance matrix of the system and the memory modes as the following,
 \begin{equation}
 \label{eq_cm_SM}
 \sigma_{\text{SM}} = \frac{1}{2}\left(
 \begin{array}{ccc}
 \sigma_{S}&\sigma_{SM_1}&\sigma_{SM_2}\\
 \sigma_{SM_1}^{T}&\sigma_{M_1}&\sigma_{M_1M_2}\\
 \sigma_{SM_2}^{T}&\sigma_{M_1M_2}^{T}&\sigma_{M_2}\\
 \end{array}
 \right),
 \end{equation}
where the subscripts $S$, $M_1$,  and $M_2$ denotes the system, $M_1$  and $M_2$ parts of the memory.
Actually, all the covariance matrices related to the von Neumann entropies in Eqs. (\ref{eq_TMI}) and (\ref{eq_BMI}) are presented in Eq. (\ref{eq_cm_SM}). It should be noticed that the dimensions of the submatrices including the memory modes, in particular the $M_2$ part, varies with the size of the memory. Additionally, once the scattering matrix is known, the covariance matrix $\sigma_{SM}$ is only determined by the initial state. The dependence of $\sigma_{SM}$ on the initial states are presented in Appendix.

\section{Results and discussions}
\label{resultsanddiscussion}
We now turn to a detailed analysis of the case with $N_m = 2$, i.e. only one mode in $M_2$ part. Under this premise, we start with the investigation on the dynamics of the local information with different initial memory and environment states to reveal the conditions under which the information scrambling occurs. Then we proceed with the discussion on the effects of the disorders and the imperfections of couplings for the scrambling process in our CM. Lastly we extend our discussion by considering the memories of different sizes.

\subsection{The initial states}
As mentioned before, the information carried by the system is locally encoded in the first particle of the memory in terms of the TMSV state at initial time. We recall that the characteristic function of the $|\text{TMSV}(\xi)\rangle$ is give by
\begin{eqnarray}
\chi^{\text{in}}_{SM_1}\left(\mu_S,\mu_{M_1}\right)&=&\exp{\left( \frac{\mu_S\mu_{M_1}+\mu_S^*\mu_{M_1}^*}{2}\sinh{\xi}\right)}\cr\cr
&& \times \exp{\left( -\frac{|\mu_S|^2+|\mu_{M_1}|^2}{2}\cosh{\xi}  \right)}.
\end{eqnarray}
Here we set the squeezing parameter $\xi$ to be real for simplicity.

We set the initial state of the $M_2$ part and the environment to be a tensor product of the uncorrelated single-mode squeezing vacuum (SMSV) state, i.e. $\rho_{M_2E}=\bigotimes_k{\hat{S}(\xi_k)|0\rangle_k\langle 0|\hat{S}^\dagger(\xi_k)}$ where $\xi_k=r_ke^{i\phi_k}$, $r_k$ and $\phi_k$ are the squeezed strength and angle for mode $k$, respectively.
The characteristic function of the $k$-th SMSV state is
\begin{equation}
\chi^{\text{in}}_k\left(\mu_k\right)=\exp{\left(-\frac{|\mu_k|^2}{2}\cosh{2r_k}+\frac{\mu_k^2e^{-i\phi_k} +(\mu_k^*)^2e^{i\phi_k} }{2}\sinh{2r_k}\right)},
\end{equation}
and, as a consequence, the input characteristic function of the whole system is given by
\begin{equation}
\chi_J^{\text{in}}(\bm{\mu})=\chi^{\text{in}}_{SM_1}(\mu_S,\mu_{M_1})\times\prod_{k}{\chi^{\text{in}}_k\left(\mu_k\right)} .
\end{equation}

\begin{figure}[h]
  \includegraphics[width=0.95\linewidth]{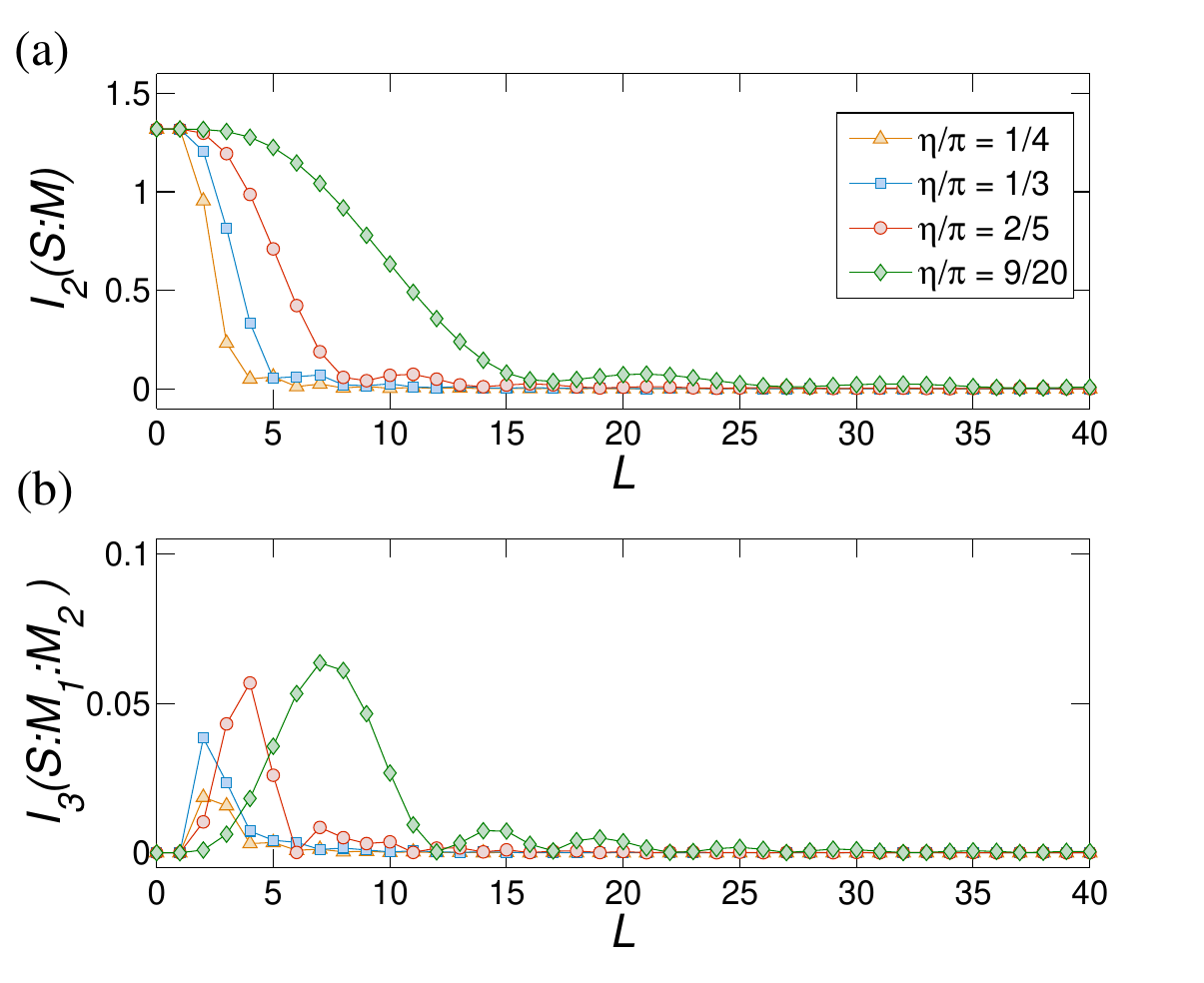}
  \caption{The $L$ dependence of BMI (a) and TMI (b) for various transmission angle $\eta$ of the BS. The input state of each mode is vacuum state. The transmission angles of the BS are $\eta=\pi/4$ (triangles), $\pi/3$ (squares), $2\pi/5$ (circles) and $9\pi/20$ (diamond), respectively.}
  \label{Fig_Transmission}
\end{figure}

Now we are ready to study the dynamics of the information in the CM. We first consider the case that all the input state of the $M_2$ and environmental modes are vacuum states, i.e. $r_k=0$ and $\phi_k=0$ for all $k$. As shown in Fig. \ref{Fig_Transmission}, the behavior of the $I_2(S:M)$, which measures the correlation between the system and memory, shows damping oscillation and decreases to zero in the long-time limit. It can be understood as the vacuum environment always absorbs the information. Moreover, the smaller transmissivity ($\eta\rightarrow\pi/2$) makes the information flow to the environment slower. It is interesting that $I_3(S:M_1:M_2)$ is always positive during the time evolution regardless of the transmissivity which is characterized by $\eta$. Namely, the information scrambling in the memory is never presented regardless of the interaction strength of the collisions when the $M_2$ and environment are vacuum states.

\begin{figure}[h]
  \includegraphics[width=0.95\linewidth]{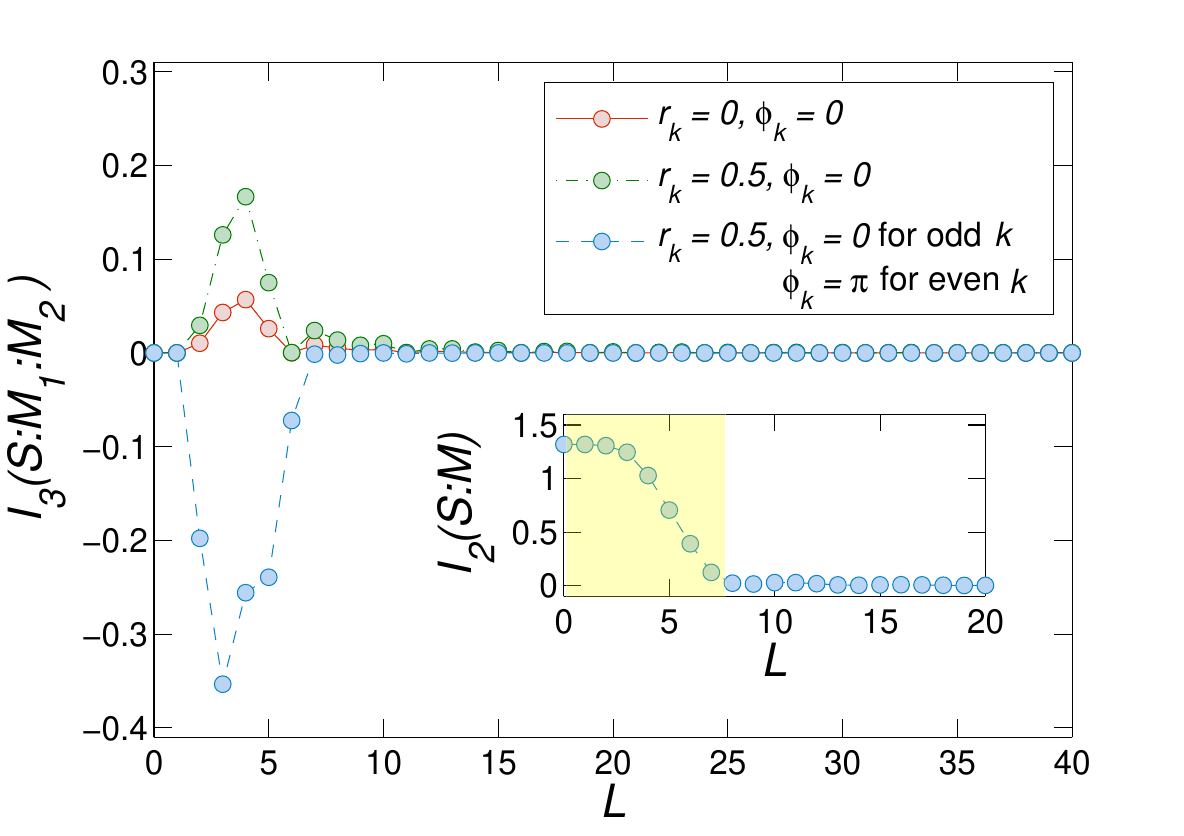}
  \caption{The $L$ dependence of TMI for the input state of each mode being vacuum state (circles connected by solid line), identical SMSV state with (i) $r_k=0.5$ and $\phi_k=0$ (circles connected by dotted-dashed line), and (ii) $r_k=0.5$ and $\phi_k=0$ for odd $k$ while $\phi_k=\pi$ for even $k$. The inset shows the $L$ dependence of BMI of case (ii). The transmission angles of the BSs are $\eta=9\pi/20$.
  }
  \label{Fig_Environment}
\end{figure}

We now investigate the influence of different types of initial states on the information scrambling. To this aim, the states are initialized by setting (i) $r_k=r\ne 0$, $\phi_k=0$ for all $k$, and (ii) $r_k=r\ne 0$, $\phi_k=0$ for the odd $k$ and $\phi_k=\pi$ for the even $k$. The former indicates that all the $M_2$ and environmental modes are squeezed with the same strength and direction, while the latter indicates that the modes are alternatively squeezed in two directions which are perpendicular to each other. In Fig. \ref{Fig_Environment}, we show the stroboscopic time-evolutions of the TMI for $\rho_{J}$ with the different initial states. For case (i), the value of $I_3(S:M_1:M_2)$ is always positive during the time evolution implying the absence of information scrambling. However, for case (ii) the value of $I_3(S:M_1:M_2)$ may become negative during the evolution, implying the presence of information scrambling. In Fig. \ref{Fig_Environment}, one can see that during the period that the information are maintaining in the memory, i.e. $I_2{(S:M)}>0$, the corresponding TMI is negative. This means that the information in the memory is scrambled and cannot be extracted by only local operations. Indeed, the BS will entangled two uncorrelated squeezed vacuum states with orthogonal squeezing directions and left uncorrelated if the two input states are squeezed in the same direction.

\subsection{The effects of disorders}
Next, we discuss the effects of disorders in the information scrambling process. The disorders in the interferometric network can be realized by introducing the phase shifts between different optical paths \cite{crespi2013a}. In our CM, the fixed phase shifters are inserted in a given optical path at each step to simulate the static disorder of the corresponding mode.

\begin{figure}[h]
  \includegraphics[width=1\linewidth]{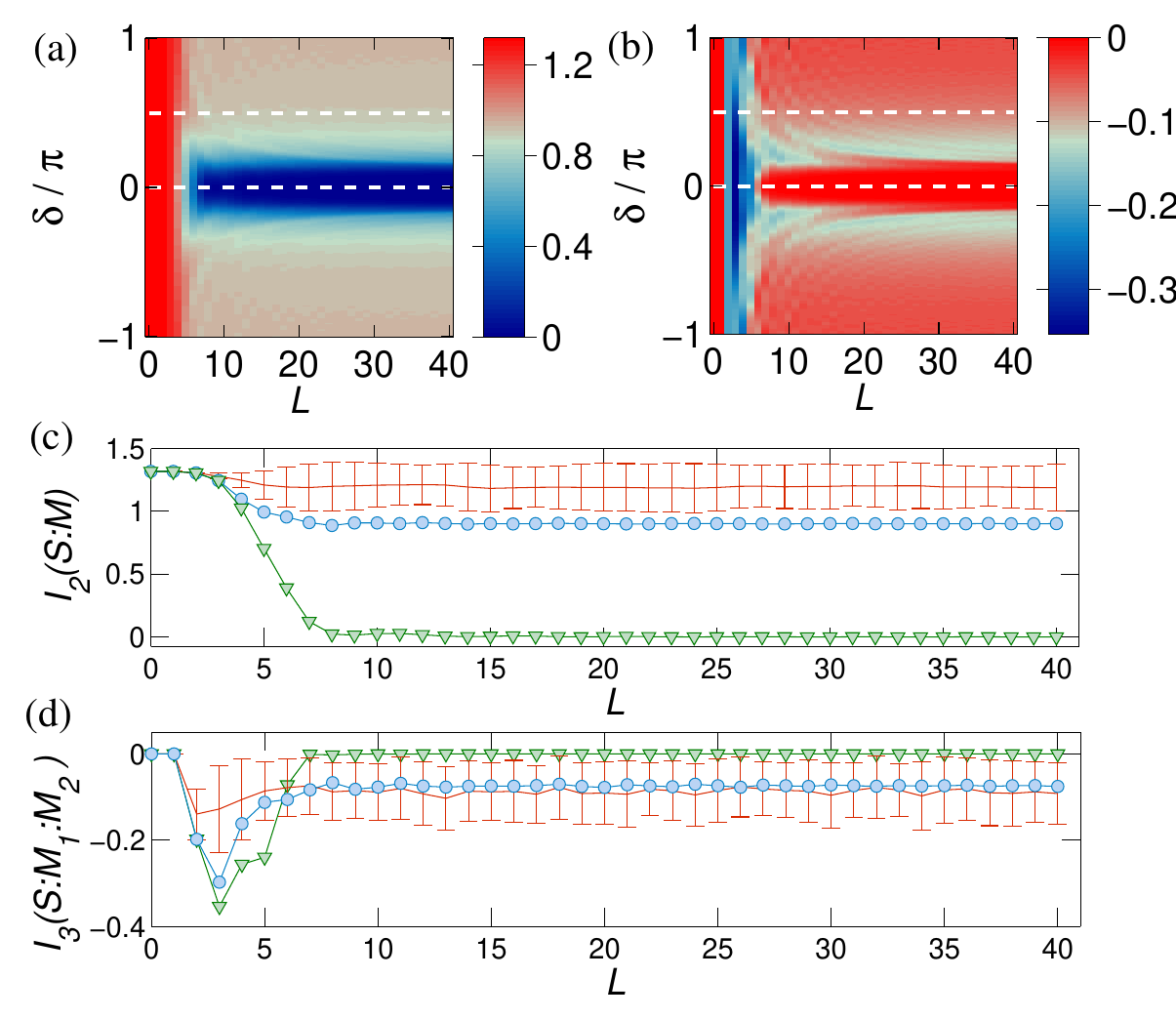}
  \caption{The $L$ dependence of BMI (a) and TMI (b) in the presence of phase difference between the $M_1$ and $M_2$ modes. The phase difference (or disorder) traps and scramble the information inside the memory. Notice the different color scales. In panels (c) and (d), the circles and triangles denotes the $L$ dependence of BMI or TMI for $\delta = 0$ and $\pi/2$ [along the dashed lines in panels (a) and (b)] in the absence of disorders in the environment. The solid lines are the $L$ dependence of the averaged BMI and TMI over 128 samples of disorders in the environment and the error bars denote the standard deviations.The transmission angles of the BSs are $\eta=9\pi/20$.}
  \label{Fig_disorder}
\end{figure}

In order to investigate the role of the disorder inside the memory, we add the phase shifters only in the $M_1$ mode at each step. This is reasonable for the memory with only two modes, because we are concerning about the relative phase difference $\delta$ of $M_1$ and $M_2$ modes. In Fig. \ref{Fig_disorder}, we show the stroboscopic dynamics of the BMI and TMI for various phase differences between the $M_1$ and $M_2$ modes. From Fig. \ref{Fig_disorder}(a) one can see that the BMI approaches to a nonzero steady-state value in the long-time limit for $\delta\ne 0$. Moreover, as as shown in Fig. \ref{Fig_disorder}(b), the TMI evolve to a negative steady-state value. For example, in Fig. \ref{Fig_disorder}(c) and (d) we show the $L$ dependence of BMI and TMI for $\delta/\pi=0$ and $\pi/2$ [along the horizontal dashed lines in panels (a) and (b)]. We can thus conclude that the presence of disorder in the memory tends to trap the information in the memory and scramble them. The disorder prevents not only the information lossing from the memory but also locally accessing the information.

We also consider the effects of the disorders in the environment. The phase shifters in the optical paths of all the memory and environment modes at each step are added to introduce the disorders $\delta$ and $\delta_k$ which denotes the disorder of the $k$-th environmental mode. The disorders $\delta_k$ are generated randomly and are static for a fixed mode. The random phases are uniformly distributed on the interval $[-\pi,\pi]$. In Fig. \ref{Fig_disorder}(c) and (d), the averaged BMI and TMI over 128 samples and their standard deviations are shown. One can see that the steady-state value of the averaged BMI is greater than that of the cases of $\delta_k=0$, i.e. disorder is present only in the memory. This means that the disorders in the environment enhance the localization of information in the memory. The residual information in the memory is scrambled as well.

\subsection{The effects of imperfect coupling}
\begin{figure}[h]
  \includegraphics[width=0.9\linewidth]{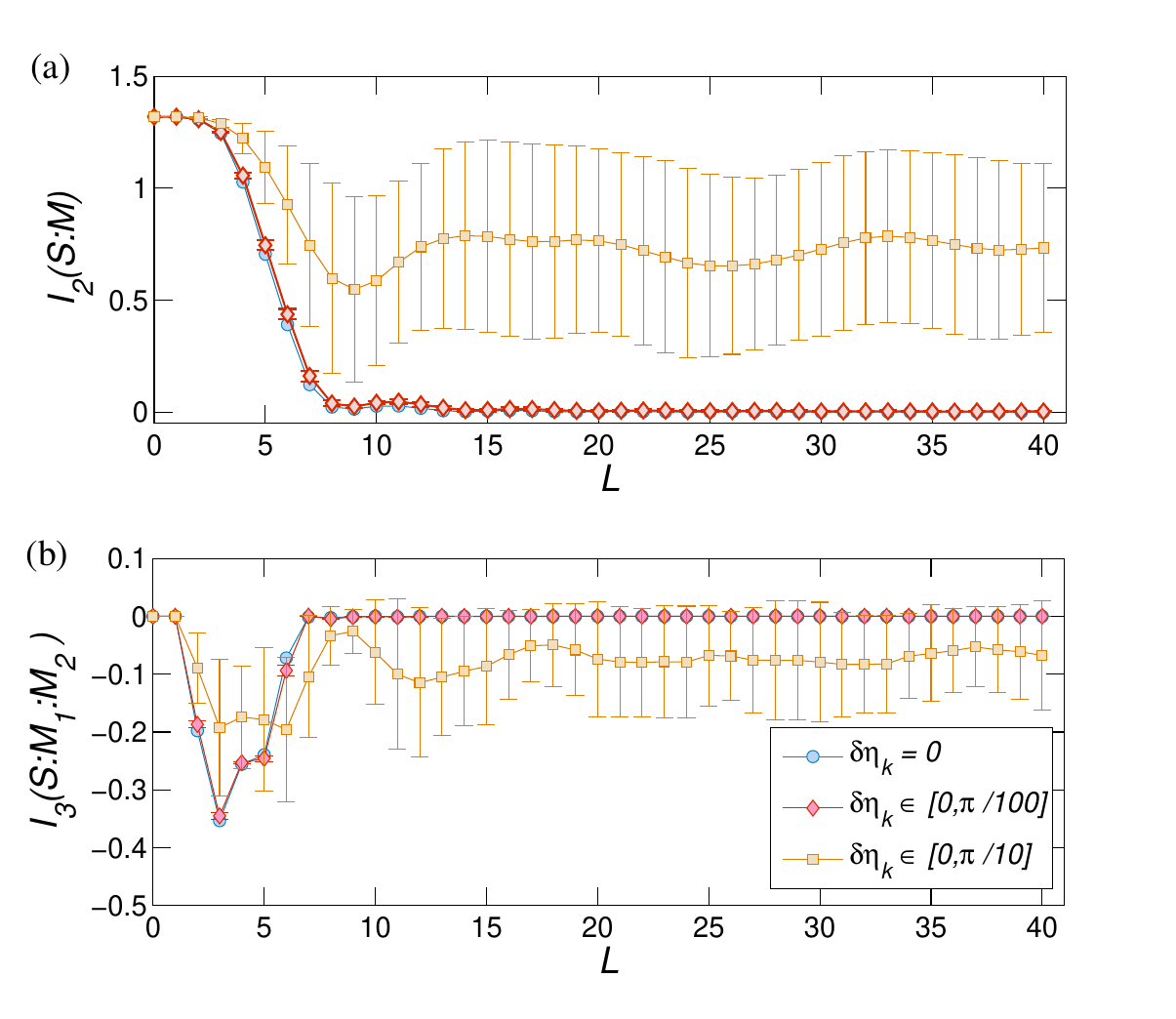}
  \caption{The stroboscopic dynamics of BMI (a) and TMI (b). The circles, diamonds and squares denote the cases of perfect, with small imperfection and large imperfection, respectively. }
  \label{Fig_imperfection}
\end{figure}

So far we have considered the case that the interaction strength in the CM are uniform which requires all the BSs in the interferometric to be perfectly identical, i.e. the transmission angles of the BSs satisfy $\eta_k=\eta$, $\forall k$. However, the perfect BSs are hard to manufactured in practice. Thus it is necessary to investigate the effects of imperfection of the BSs. To this end, we add the static imperfection to each BS for mixing neighbouring modes. The imperfections are simulated by a randomly generated fluctuation $\delta\eta_k$ around the ideal $\eta_k$. The effects of the imperfections are characterized by the averaged BMI and TMI over 128 samples. In Fig. \ref{Fig_imperfection} we show the $L$ dependence of the averaged BMI and TMI for the cases of small ($\delta\eta_k\in[0,\pi/100]$) and large ($\delta\eta_k\in[0,\pi/10]$) imperfections. The random imperfections are uniformly distributed. Compared with the perfect case ($\delta\eta_k=0$) the small imperfection changes the stroboscopic dynamics of both BMI and TMI slightly and in the long-time limit the information completely flow away from the memory which is indicated by the vanishing BMI. However, for the case of large imperfections, the averaged BMI keeps non-zero even if in the long-time limit. Furthermore the averaged TMI is negative in the long-time limit meaning that the residual information in the memory is scrambled.

\subsection{The memory size}
\begin{figure}[h]
  \includegraphics[width=0.95\linewidth]{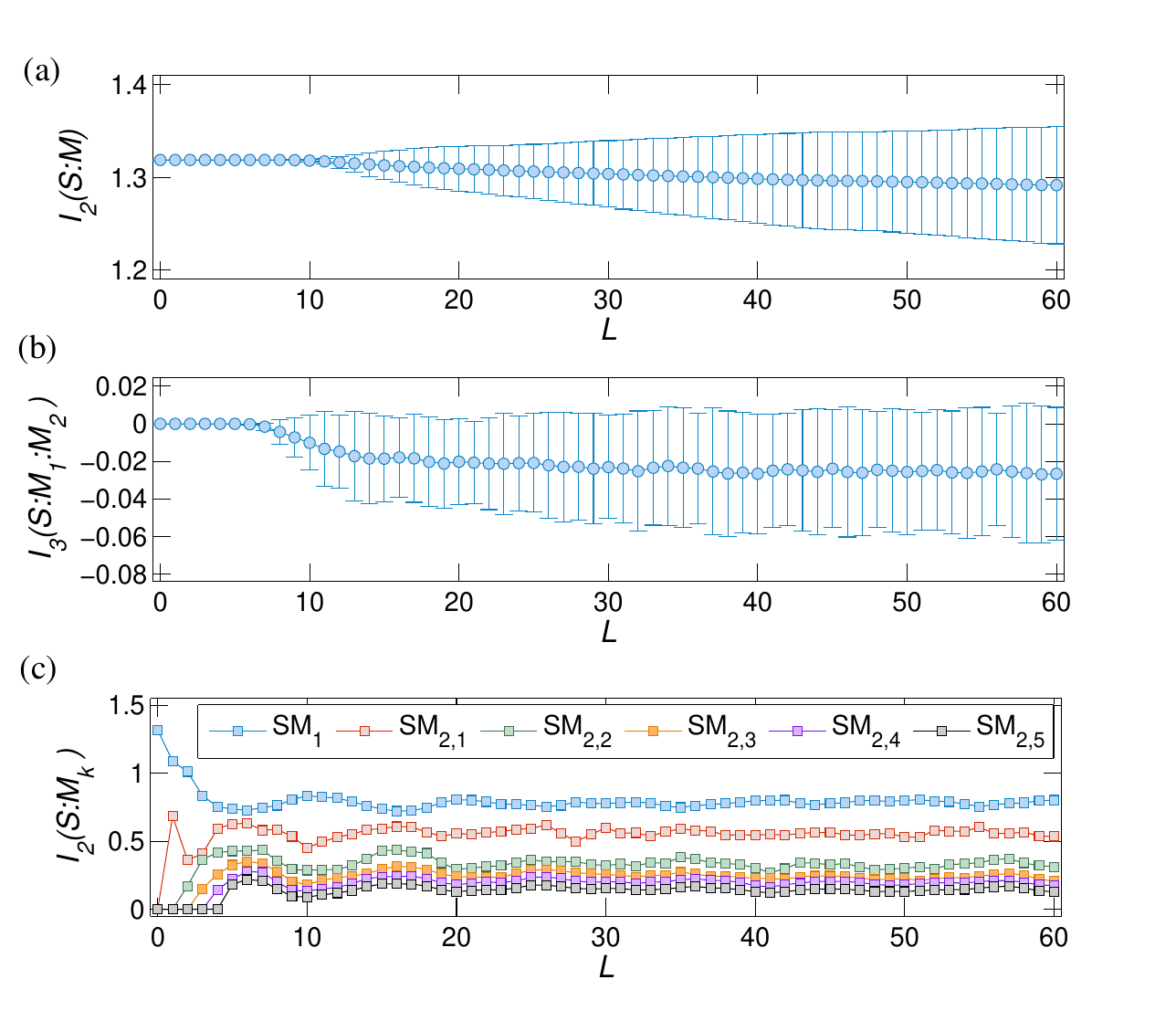}
  \caption{The stroboscopic evolution of BMI (a) and TMI (b) between the system and the memory which consists of $k=6$ modes. (c) The stroboscopic evolution of the BMIs between system and each memory mode. Note that the BMIs between system and $M_{2,k}$ modes have been triple-amplified to facilitate the observation.
   }
  \label{Fig_memorysize}
\end{figure}

Now let us consider the case that the memory consists $N_m > 2$ modes. We label the first mode as $M_1$ and the modes left as $M_{2,k}$ ($2\le k \le N_m$).
Suppose that initially the system and $M_1$ modes are in the TMSV state and the states of the $M_{2,k}$ modes are in the SMSV state with alternative squeezing angles ($\phi_k = 0$ for odd $k$ and $\phi_k=\pi$ for even $k$). In Fig. \ref{Fig_memorysize}, we show the averaged stroboscopic evolutions of the BMI and TMI of the system and memory in the presence of disorders in the memory. One can see that the BMI between the system and memory remains close to the initial value even in the long-time limit due to the information localization by the disorder. However, the TMI among the system, $M_1$ and $M_2$ modes approaches to a stationary negative value meaning that the information inside the memory is scrambled. In order to trace the information initially encoded in the $M_1$ mode, we compute the BMIs between the system and any memory modes in Fig. \ref{Fig_memorysize}(c). As the time pasts, we see that the correlation between the system and $M_1$ mode decreases while the correlations between system and other modes in $M_2$ are built. This means that the initial locally encoded information delocalizes through the whole memory and is separately stored in each mode of the memory.

\section{Summary}
\label{summary}
As a summary, we have investigated the information scrambling of Gaussian states in an interferometric network based on the CM. We have studied the stroboscopic evolution of the BMI and TMI of the system $S$ and the memory $M$. The information of the system is initially locally encoded in the first unit of the memory in terms of an entangled TMSV state. We have found that the information will eventually lost from the memory for the vacuum or squeezed vacuum environment. The decay rate is determined by the transmission angle of the BSs. However, if the environment units are prepared in the squeezed vacuum states along two directions perpendicular to each other, the information scrambling appears before the information completely lost from the memory. This phenomenon is similar to that the XXX spin chain in N{\' e}el state has the possibility to scramble quantum information while never scramble information in spin-all-up state \cite{iyoda2018}.

We have also investigated the effects of disorders on the stroboscopic evolution of the BMI and TMI. It is interesting that the disorders trap the information inside the memory revealing by the nonzero BMI in the long-time limit. The trapped information is scrambled which is manifested by the negative TMI. This means that the disorders plays positive roles in preventing the information lost from the memory and locally accessing the information. Additionally we have investigated the effects of imperfections of the BSs, we found that the small imperfection does not change the dynamics of the information while large imperfection tends to localize the information. Additional simulations with more samples of both disorders and imperfections confirm the validity of the present results \cite{convergency}.

We have considered the case that the memory consists of more than two modes in the presence of disorders in the memory. In such a case, we find that the initial information is trapped inside the memory and scrambled. This is supported by the fact that the correlations between each memory modes are built at the cost of the initial correlations between the system and $M_1$ modes.

Finally, we would like to give a brief discussion on the possible the experimental realization of our CM. A mature and promising candidate to implement our scheme is the advanced integrated photonic quantum simulators \cite{crespi2013a,crespi2013b}. This platform has the advantages of integrability, tunability and high efficiency of detection \cite{AAG2012,wang2019}. It has been utilized to simulate the Anderson localization \cite{crespi2013a} and Boson sampling \cite{latmiral2016,paesani2019,Huiwang2019}. In particular, the studies on the Gaussian Boson sampling with linear optics elements from both the theoretical \cite{hamilton2017,quesada2018,kruse2019} and experimental \cite{zhong2019} perspectives support the experimental realization of our CM.

\acknowledgments
We thank Rosario Fazio and Silvia Pappalardi for fruitful discussion.
This work is supported by National Natural Science Foundation of China under
Grant No. 11975064 and No. 11775040, and the Fundamental Research Funds for the Central Universities No. DUT19LK13.

\section*{APPENDIX}
\label{appendix}
In this appendix, we show the specific forms of Eq.(\ref{eq_cm_SM}) with different given initial states. Before the detailed discussions are made, it is important to note that, the parameters of the input state $\rho^{in}_{J}$ we set are identical with the main context, which can be divide into three cases:

(1) $r_k=0$ and $\phi_k=0$, for considering all the input states of the $M_2$ and the environmental modes as vacuum states;

(2) $r_k = r\ne 0 $ and $\phi_k=0$, for the case of the initial states are all be consider as the SMSV state, in which the modes have the same strength and direction;

(3) $r_k=r\ne0$, $\phi_k=0$ for the even $k$ and $\phi_k = \pi$ for the odd $k$, in this case, the initial states are also be set as SMSV state, but the squeezed direction of each neighboring mode is perpendicular.

The matrix elements $\sigma_{S}$, $\sigma_{SM_1}$ and $\sigma_{M_1}$ of the covariance matrix in Eq.(\ref{eq_cm_SM}) are two-dimensional matrices given as the following,

\begin{equation}
 \label{eq_cm_SM_S}
 \sigma_{\text{S}} = \left(
 \begin{array}{cc}
 \cosh(2\xi)&0\\
 0&\cosh(2\xi)\\
 \end{array}
 \right),
 \end{equation}
\begin{equation}
\label{eq_cm_SM_SM1}
\sigma_{\text{SM}_{1}} = \left(
\begin{array}{cc}
\sinh(2\xi)\Re(S_{2,2}^{*})&\sinh(2\xi)\Im(S_{2,2}^{*})\\
\sinh(2\xi)\Im(S_{2,2}^{*})&-\sinh(2\xi)\Re(S_{2,2}^{*})\\
\end{array}
\right),
\end{equation}
\begin{equation}
 \label{eq_cm_SM_M1}
 \sigma_{M_1} = \left(
 \begin{array}{cc}
 \alpha_{M_1}+\beta_{M_1}&\gamma_{M_1}\\
 \gamma_{M_1}&-\alpha_{M_1}+\beta_{M_1}\\
 \end{array}
 \right),
 \end{equation}
with the corresponding elements,
\begin{equation}
\begin{aligned}
\alpha_{\text{M}_{1}}&=\frac{1}{2}\sinh(2r_k) \sum _{k=3} ^{L+2} (e^{i\phi_k}S_{k,2}^{*2} + e^{-i\phi_k}S_{k,2}^{2}),\\
\beta_{\text{M}_{1}}&=\cosh(2\xi)\vert S_{2,2} \vert ^{2}+\cosh(2r_k)(1-\vert S_{2,2} \vert ^{2}),\\
\gamma_{\text{M}_{1}}&=\frac{1}{2i}\sinh(2r_k) \sum_{k=3} ^{L+2} \left(e^{i\phi_k}S_{k,2}^{*2} - e^{-i\phi_k}S_{k,2}^{2}\right),\\
\end{aligned}\nonumber
\end{equation}
where $S_{i,j}$ is the matrix element of the scattering matrix $\mathbb{S}^{-1}(L)$ which is given in Eq.(\ref{SL}).

However, the dimensions of the matrices $\sigma_{M_2}$, $\sigma_{SM_{2}}$ and $\sigma_{M_{1}M_{2}}$ are increasing with the size of the memory, and a general form of those matrices can be obtained by,
\begin{equation}
\label{eq_cm_SM_M2}
\sigma_{M_2}=
\begin{pmatrix}
\sigma_{11}&\sigma_{12}&\cdots&\sigma_{1N_m}\\
\sigma_{21}&\sigma_{22}&\cdots&\sigma_{2N_m}\\
\vdots&\vdots&\ddots&\cdots\\
\sigma_{N_m1}&\sigma_{N_m2}&\cdots&\sigma_{N_{m}N_{m}}\\
\end{pmatrix},
\end{equation}
with a fixed form of
\begin{equation}
\sigma_{aa} =\left(
\begin{array}{cc}
\alpha_{M_2}+\beta_{M_2}&\gamma_{M_2}\\
\gamma_{M_2}&-\alpha_{M_2}+\beta_{M_2}\\
\end{array}
\right),
\label{sigmaaa}
\end{equation}
\begin{equation}
\sigma_{ab} = \left(
\begin{array}{cc}
\mathcal{A}_{M_2}+\mathcal{B}_{M_2}&\mathcal{C}_{M_2}+\mathcal{D}_{M_2}\\
\mathcal{C}_{M_2}-\mathcal{D}_{M_2}&-\mathcal{A}_{M_2}+\mathcal{B}_{M_2}\\
\end{array}
\right),
\label{sigmaab}
\end{equation}
the elements in Eq.(\ref{sigmaaa}) and Eq.(\ref{sigmaab}) are
\begin{equation}
\begin{aligned}
\alpha_{M_2}&=\frac{1}{2}\sinh(2r_k)\sum_{k=3}^{L+2}(e^{i\phi_{k}}S_{k,a+2}^{*2}+e^{-i\phi_{k}}S_{k,a+2}^{2}),\\
\beta_{M_2}&=\cosh(2\xi)\vert S_{2,a+2} \vert ^{2} + \cosh(2r_k)(1-\vert S_{2,a+2}\vert ^{2}), \\
\gamma_{M_2} &= \frac{1}{2i}\sinh(2r_k)\sum_{k=3}^{L+2}\left(e^{i\phi_{k}}S_{k,a+2}^{*2} - e^{-i\phi_{k}}S_{k,a+2}^{2}\right),\\
\end{aligned}\nonumber
\end{equation}
and
\begin{equation}
\begin{aligned}
\mathcal{A}_{M_2} &= \frac{1}{2} \sinh(2r_k) \sum_{k=3}^{L+2} \left(e^{i\phi_k} S_{k,a+2}^{*} S_{k,b+2} + e^{-i\phi_k} S_{k,a+2} S_{k,b+2}^{*} \right),\\
\mathcal{B}_{M_2} &= \cosh(2\xi) \Re (S_{2,a+2} S_{2,b+2}^{*}) + \cosh(2r_k) \Re(- S_{2,a+2} S_{2,b+2}^{*}),\\
\mathcal{C}_{M_2} &= \frac{1}{2i} \sinh(2r_k) \sum_{k=3}^{L+2} \left(e^{i\phi_k} S_{k,a+2}^{*} S_{k,b+2} - e^{-i\phi_k} S_{k,a+2} S_{k,b+2}^{*} \right),\\
\mathcal{D}_{M_2} &= \cosh(2\xi) \Im (S_{2,a+2} S_{2,b+2}^{*}) + \cosh(2r_k) \Im(- S_{2,a+2} S_{2,b+2}^{*}),\\
\end{aligned}\nonumber
\end{equation}
the rest two matrices, $\sigma_{SM_{2}}$ and $\sigma_{M_{1}M_{2}}$ can be derived by
\begin{equation}
\label{eq_cm_SM_SM2}
\sigma_{SM_{2}} =
\begin{pmatrix}
\sigma_{S}&\sigma_{SM_{2,1}}&\cdots&\sigma_{SM_{2,N_{m}}}\\
\sigma_{M_{2,1}S}&\sigma_{11}&\cdots&\sigma_{1N_{m}}\\
\vdots&\vdots&\ddots&\cdots\\
\sigma_{M_{2,N{m}}S}&\sigma_{N_{m}1}&\cdots&\sigma_{N_{m}N_{m}}\\
\end{pmatrix},
\end{equation}
\begin{equation}
\sigma_{\text{SM}_{2,a}} = \left(
\begin{array}{cc}
\sinh(2\xi)\Re(S_{2,a+2}^{*})&\sinh(2\xi)\Im(S_{2,a+2}^{*})\\
\sinh(2\xi)\Im(S_{2,a+2}^{*})&-\sinh(2\xi)\Re(S_{2,a+2}^{*})\\
\end{array}
\right),
\end{equation}
and

\begin{equation}
\label{eq_cm_SM_M1M2}
\sigma_{M_{1}M_{2}} =
\begin{pmatrix}
\sigma_{M_1}&\sigma_{M_{1}M_{2,1}}&\cdots&\sigma_{M_{1}M_{2,N_{m}}}\\
\sigma_{M_{2,1}M_{1}}&\sigma_{11}&\cdots&\sigma_{1N_{m}}\\
\vdots&\vdots&\ddots&\cdots\\
\sigma_{M_{2,N{m}}M_{1}}&\sigma_{N_{m}1}&\cdots&\sigma_{N_{m}N_{m}}\\
\end{pmatrix},
\end{equation}
\begin{equation}
\sigma_{M_{1}M_{2,a}} = \left(
\begin{array}{cc}
\mathcal{A}_{M_{1}M_{2}}+\mathcal{B}_{M_{1}M_{2}}&\mathcal{C}_{M_{1}M_{2}}+\mathcal{D}_{M_{1}M_{2}}\\
\mathcal{C}_{M_{1}M_{2}}-\mathcal{D}_{M_{1}M_{2}}&-\mathcal{A}_{M_{1}M_{2}}+\mathcal{B}_{M_{1}M_{2}}\\
\end{array}
\right),
\end{equation}
with the matrix elements being
\begin{equation}
\begin{aligned}
\mathcal{A}_{M_{1}M_{2}} &= \frac{1}{2} \sinh(2r_k) \sum_{k=3}^{L+2} \left( e^{i\phi_k} S_{k,2}^{*} S_{k,a+2} + e^{-i\phi_k} S_{k,2} S_{k,a+2}^{*} \right),\\
\mathcal{B}_{M_{1}M_{2}} &= \cosh(2\xi) \Re (S_{2,2} S_{2,a+2}^{*}) + \cosh(2r_k) \Re(- S_{2,2} S_{2,a+2}^{*}),\\
\mathcal{C}_{M_{1}M_{2}} &= \frac{1}{2i} \sinh(2r_k) \sum_{k=3}^{L+2} \left( e^{i\phi_k} S_{k,2}^{*} S_{k,a+2} - e^{-i\phi_k} S_{k,2} S_{k,a+2}^{*} \right),\\
\mathcal{D}_{M_{1}M_{2}} &= \cosh(2\xi) \Im (S_{2,2} S_{2,a+2}^{*}) + \cosh(2r_k) \Im(- S_{2,2} S_{2,a+2}^{*}).\\
\end{aligned}\nonumber
\end{equation}


\begin{thebibliography}{999}
\bibitem{convergency}
We also checked the dynamics of TMI and BMI averaged over 256, 512 and 1024 samples in the presence of disorders, compared with Figs. \ref{Fig_disorder}(c) and (d), and imperfections, compared with Fig. \ref{Fig_imperfection}. The results show that the values of BMI and TMI averaging over 128 samples have already converged.
\end{thebibliography}
\end{document}